\documentclass[12pt]{article}
\usepackage{graphics,epsfig}
\usepackage[english]{babel}
\newcommand{\cinst}[2]{$^{\mathrm{#1}}$~#2\par}
\newcommand{\crefi}[1]{$^{\mathrm{#1}}$}

\begin{document}


\thispagestyle{empty}
\begingroup


\vglue.5cm \hspace{9cm}{YerPhI Preprint - 1626 (2013)
\vglue.5cm
\begin{center}

{\normalsize \bf STUDY OF NUCLEAR EFFECTS IN THE INCLUSIVE
\\[.3cm]
NEUTRINOPRODUCTION OF $\Delta^{++}(1232)$} 

\end{center}

\vspace{1.cm}
\begin{center}

 N.M.~Agababyan\crefi{1},
 N.~Grigoryan\crefi{2}, H.~Gulkanyan\crefi{2},\\
 A.A.~Ivanilov\crefi{3}, V.A.~Korotkov\crefi{3}

\setlength{\parskip}{0mm} \small

\vspace{1.cm} \cinst{1}{Joint Institute for Nuclear Research,
Dubna, Russia} \cinst{2}{Alikhanyan National Scientific Laboratory \\
(Yerevan Physics Institute), Armenia}
\cinst{3}{Institute for High Energy Physics, Protvino, Russia}
\end{center}

\setlength{\parskip}{0mm}
\small

\vspace{130mm}

{\centerline{\bf YEREVAN  2013}}

\newpage
\vspace{1.cm}

\begin{abstract}
For the first time, the inclusive neutrinoproduction of $\Delta^{++}(1232)$ from nuclei is investigated. The total yield of $\Delta^{++}(1232)$ in neutrino-nuclear interaction is found to be $\langle n\rangle_{\nu A} = 0.080\pm0.011$, being compatible with  $\langle n\rangle_{\nu N} = 0.091\pm0.013$ inferred for neutrino-nucleon interactions. It is shown that the $\Delta^{++}(1232)$ inclusive spectra on variables $x_F$, $z$ and the isobar total momentum for $\nu A$ interactions are shifted toward lower values respective to those for $\nu N$ interactions, that can be caused by secondary intranuclear collision processes. No idication on a significant nuclear absorption of $\Delta^{++}(1232)$ leading to its disappearance is found.
\end{abstract}

\newpage
\setcounter{page}{1}
\begin{center}
{\large 1. ~INTRODUCTION}\\
\end{center}

Experimental investigations of the leptoproduction of hadronic resonances on nuclei are necessary for a deeper insight into the space-time structure of the quark string fragmentation and the formation of hadrons, a significant fraction of which originates from the resonance decays (see \cite{ref1} and references therein). At present, the experimental data concerning the nuclear medium influence on the inclusive neutrinoproduction of resonances are rather scarce and concern only mesonic resonances (see e.g. \cite{ref2,ref3} concerning $\rho$ mesons), while no data concerning baryonic resonances are available.

The aim of this work is to obtain the first experimental data on the $\Delta^{++}(1232)$ inclusive neutrinoproduction from nuclear targets, providing an information about nuclear effects in this process. In Section 2, the experimental procedure is described. The experimental results are presented in Section 3 and summarized in Section4. 

\begin{center}
{\large 2. ~EXPERIMENTAL PROCEDURE}\\
\end{center}

The experiment was performed with SKAT bubble chamber \cite{ref4},
exposed to a wideband neutrino beam obtained with a 70 GeV primary
protons from the Serpukhov accelerator. The chamber was filled
with a propane-freon mixture containing 87 vol\% propane
($C_3H_8$) and 13 vol\% freon ($CF_3Br$) with the percentage of
nuclei H:C:F:Br = 67.9:26.8:4.0:1.3 \%. A 20 kG uniform magnetic
field was provided within the operating chamber volume.

Charged-current interactions containing a negative muon with
momentum $p_{\mu} >$0.5 GeV/c were selected. The overwhelming part of
protons with momentum below 0.6 GeV$/c$ and a fraction of protons  with momentum
0.6-0.85 GeV$/c$ were identified by their stopping in the chamber.
Stopping $\pi^+$ mesons were identified by their
$\pi^+$-$\mu^+$-$e^+$ decay. A fraction of low-momentum
($p_{\pi^+} < 0.5$ GeV$/c$) $\pi^+$ mesons were identified by the
mass-dependent fit provided that the $\chi^2$- value for the pion
hypothesis was significantly smaller as compared to that for
proton. Non-identified positively charged hadrons are assigned the pion mass or,
in the cases explained below, the proton mass.
It was required the errors in measuring the momenta be
less than 24\% for muon, 60\% for other charged particles and
$V^0$'s (corresponding to neutral strange particles) and less than 100\%
for photons. The mean relative error ($\Delta p/p$) in the momentum measurement for
muons, pions, protons and gammas was, respectively, 3\%, 6.5\%, 10\% and 19\%.
Each event was given a weight to correct for the
fraction of events excluded due to improperly reconstruction. More
details concerning the experimental procedure, in particular, the
reconstruction of the neutrino energy $E_\nu$, can be found in our
previous publications \cite{ref5,ref6}.

The events with $3< E_{\nu} <$ 30 GeV were accepted. 
The total number of accepted events was 8390, with 
the mean value of the neutrino energy  
$\langle E_\nu\rangle$ = 8.7 GeV.
Below we will study (and compare) the inclusive $\Delta^{++}(1232)$ production in two subsamples of events: the quasinucleon subsample ($B_N$) and nuclear subsample ($B_A$) which are composed as described below (see for details \cite{ref6,ref7}).  
The  subsample $B_N$ includes the events without any indication of the nuclear disintegration or a secondary intranuclear interaction,
satisfying the following topological and kinematic criteria: the net charge of secondary hadrons was equal to +1 (for quasineutron subsample $B_n$) or +2 (for quasiproton subsample $B_p$); the number of recorded baryons (these included
identified protons and $\Lambda$ hyperons, along with neutrons
that suffered a secondary interaction in the chamber) was
forbidden to exceed unity, baryons flying in the backward
hemisphere being required to be absent among them; the effective
target mass $M_t <  $ 1.2 GeV/$c^2$, the $M_t$ being defined as
$M_t = \sum{(E_i - p_i^L)}$ where the summation was performed over
the energies $E_i$ and the longitudinal momenta $p_i^L$ (along the
neutrino direction) of all recorded secondary particles. As it was shown in detail
in \cite{ref7,ref8}, the multiplicity and spectral characteristics of charged hadrons in the $B_p$ and $B_n$ subsamples were quite compatible with the available data obtained on hydrogen and deuterium targets.

Events which did not satisfy the aforementioned criteria were included in the 
subsample $B_S$ of 'cascade' events. 
The numbers of accepted events of subsamples  $B_p$, $B_n$ and $B_S$ were equal to
1839, 2393 and 4158, respectively. Finally, we excluded from the $B_p$ subsample
the events-candidates to the exclusive reaction $\nu p \rightarrow \mu^- p
\pi^+$ (305 events, see for details \cite{ref9,ref10}).
As it follows from the composition of the propane-freon mixture (see above), 36.7\% of the subsample $B_p$ is contributed by interactions with free hydrogen. Weighting the quasiproton events with a factor of
0.633, one can exclude the hydrogen contribution and compose 
a quasinucleon subsample $B_N$ = $B_n$ + 0.633$B_p$, corresponding to the 
neutrino interactions with peripheral nucleons of the target nuclei
(named $\nu N$ interactions), and a 'pure' nuclear subsample 
$B_A$ = $B_S$ + $B_n$ + 0.633$B_p$, corresponding to the 
neutrino interactions with nucleons of the target nuclei
(named $\nu A$ interactions).
The effective atomic weight for the nuclear subsample $B_A$ was estimated to
be approximately equal to $A_{eff}$ = 21 \cite{ref11}.

Only in a small fraction of quasinucleon
events (23.7\% in $\nu p$ and 16.3\% in $\nu n$ interactions) an
identified proton ($n_p^{id} = 1$) was present. In the remaining
events (with $n_p^{id}$ = 0) the proton hypothesis was applied to
a non-identified positively charged hadron (if any)  provided that
the proton hypothesis was not rejected by the momentum-range
relation in the propane-freon mixture. This hypothetical proton,
after introduction of a proper correction for its momentum, was
combined with an accompanying positively charged hadron (an
identified $\pi^+$ or a non-identified hadron) to compose a
hypothetical $\pi^+ p$ combination. The most part of such
combinations, especially in events containing two or more
unidentified hadrons ($n_h^{nid} \geq 2$), is expected to be
spurious due to the proton misidentification. As a result, the
angular distribution of a 'proton' in the pion-'proton' rest frame
turns out to be strongly shifted towards the negative values of
$\cos\vartheta^*_p$ , where $\vartheta^*_p$ is the angle between
the 'proton' direction and the direction of the Lorentz boost from
the lab system to the pion-'proton' rest system. As it has been
shown by simulations \cite{ref10}, 
the  $\cos\vartheta^*_p$ distribution for
the case of spurious $\pi^+ p$ combinations is strongly peaked at
$\cos\vartheta^*_p \approx -1$ and rapidly falls with increasing
$\cos\vartheta^*_p$ up to $\cos\vartheta^*_p \sim -0.6$, then
begins to flatten and becomes almost uniform at $\cos\vartheta^*_p
>0$. In order to reduce the share of spurious combinations in the
experimental $\pi^+ p$ effective mass distribution for events with
$n_p^{id} = 0$ and $n_h^{nid} \geq 2$, a cut $\cos\vartheta^*_p >
-0.6$ was applied for the combinations of two unidentified
hadrons, while those with $-0.6 < \cos\vartheta^*_p < 0$ a weight
were ascribed, so that the total numbers of combinations with
$\cos\vartheta^*_p < 0$ and $\cos\vartheta^*_p > 0$ (including
those for events with $n_p^{id} = 1$ or $n_h^{nid} < 2$) turned
out to be equal in the $\Delta^{++}(1236)$ peak region of $1.16 <
m_{\pi^+ p} < 1.32$ GeV$/c^2$. This procedure, as it was shown in \cite{ref10},
enables to improve the signal to background ratio in the $\Delta^{++}(1236)$ peak region and somewhat decreases the errors in the determination of its yield.
A similar weighting procedure was applied for those events of the subsample $B_S$ which contain two or more track-candidates to the proton (composing about 75\% of 
the subsample $B_S$). 

\begin{center}
{\large 3. ~EXPERIMENTAL RESULTS}\\
\end{center}

Two different methods were applied to infer the yield of $\Delta^{++}(1232)$ in $\nu N$ interactions. In the first method, a properly weighted combination of the recently measured yields \cite{ref10} in $\nu p$ and $\nu n$ interactions, $\langle n\rangle_{\nu p}$ and $\langle n\rangle_{\nu
n}$ , was used. The resulted total yield, $\langle n\rangle_{\nu N}$, and the integrated yields in two ranges of the Feynman $x_F$ variable, $x_F< 0$ and $x_F > 0$, are presented in Table 1, where the data from \cite{ref10} are also shown.
\noindent
\begin{table}[ht]
\caption{The $\Delta^{++}(1232)$ yields in 
$\nu p$, $\nu n$, $\nu N$, $\nu A$ and cascade-type interactions at three different ranges of $x_F$.}
\begin{center}

\begin{tabular}{|l|c|c|c|c|c|}
  \hline

&$\nu p$&$\nu n$&$\nu N$&cascade& $\nu A$ \\ 
&&&(combined)&subsample&(combined)
\\ \hline
all $x_F$& 0.170$\pm$0.029&0.051$\pm$0.012&0.087$\pm$0.012&0.080$\pm$0.022&0.083$\pm$0.013 \\
$x_F < 0$& 0.101$\pm$0.023&0.043$\pm$0.011&0.061$\pm$0.010&0.064$\pm$0.020&0.062$\pm$0.012 \\
$x_F > 0$& 0.057$\pm$0.018&0.010$\pm$0.009&0.024$\pm$0.008&0.028$\pm$0.009&0.026$\pm$0.006 \\ \hline

\end{tabular}
\end{center}
\end{table}

In the second method, the $\Delta^{++}(1232)$ yield was inferred directly from the
$\pi^+ p$ effective mass distributions for the $B_N$ subsample plotted in Figure 1. 
Following \cite{ref10} (see also \cite{ref12}), the distributions are fitted by a five-parameter function
\begin{equation}
F(m) = C_1 \cdot BW(m) + C_2 \cdot BG(m) \,  ,
\end{equation}
\noindent where $BW(m)$ is the relativistic Breit-Wigner function \cite{ref13}, smeared according to
the experimental resolution, while the background distribution was parametrized as

\begin{equation}
BG(m) = q^\alpha \cdot \exp(-\beta m^\gamma) \, ,
\end{equation}
\noindent where $q$ was the pion momentum in the $\pi^+p$ rest frame. The results of the fit (with free
parameters $C_1, C_2$, $\alpha, \beta$, $\gamma$) are shown in Figure 1 and Table 2.
\begin{table}[ht]
\caption{The $\Delta^{++}(1232)$ yields in $\nu N$, and $\nu A$ interactions at three different ranges
of $x_F$.}
\begin{center}
\begin{tabular}{|l|c|c|}
  \hline

&$\nu N$&$\nu A$
\\ \hline
all $x_F$& 0.091$\pm$0.013&0.080$\pm$0.014 \\
$x_F < 0$& 0.059$\pm$0.010&0.058$\pm$0.013 \\
$x_F > 0$& 0.028$\pm$0.009&0.029$\pm$0.006   \\ \hline

\end{tabular}
\end{center}
\end{table}
 
Both methods were also applied to infer the  $\Delta^{++}(1232)$ yields in $\nu A$ 
interactions. In the first method, the $\langle n\rangle_{\nu A}$ was evaluated as a properly weighted combination of yields in $\nu p$, $\nu n$ and cascade-type interactions quoted in Table 1, while in the second method $\langle n\rangle_{\nu A}$ was inferred directly from the $\pi^ + p$ effective mass distributions for the $B_A$ subsample plotted in Figure 1. The results of the fit by the function (1) are shown in Figure 1 and Table 2.

As it follows from Tables 1 and 2, both methods lead to practically the same results. Moreover, our data on the total and integrated over $x_F <0$ and $x_F >0$ regions yields do not reveal valuable nuclear effects. This statements is not, however, the case for differential yields presented below. 

In the rest part of this paper, we present the results obtained with the help of the second method for the yield estimation. Figures 2 and 3 show the inclusive spectra of  $\Delta^{++}(1232)$ on the $x_F$ variable, the squared transverse momentum $p_T^2$ variable ($p_T$ being defined respective to the intermediate $W$ boson direction), the isobar total momentum $p$ variable and the $z$ variable $-$ the fraction of the intermediate boson energy $\nu = E_{\nu} - E_{\mu}$ transferred to the $\Delta^{++}(1232)$ and defined as $z = (E-m_p)/\nu$, where $E$ is the isobar total energy and $m_p$ is the proton mass. As it is seen, the distributions on $x_F$, $z$ and $p$ in $\nu A$ interactions are noticeably shifted towards lower values respective to $\nu N$ interactions. This shifting can be caused by the $\Delta^{++}(1232)$ energy and longitudinal momentum losses in the nuclear medium due to both inelastic and elastic scattering on intranuclear nucleons which, according to theoretical predictions \cite{ref14,ref15,ref16,ref17,ref18}, might occur with cross sections compatible with (or even exceeding) the typical nucleon-nucleon cross sections. It should be noted, however, that our data on the $p_T^2$-distributions (Figure 2) do not reveal significant nuclear effects: for both $\nu N$ and $\nu A$ interactions, the $p_T^2$ spectrum can be described by an exponential function with the same slope parameter $b(\nu N) = 4.2\pm0.7$ and $b(\nu A) = 4.3\pm1.1$ (GeV$/c)^{-2}$, respectively.

Further, the data on the $\Delta^{++}(1232)$ momentum distribution (plotted in Figure 3) indicate that the intranuclear absorption of $\Delta^{++}(1232)$ leading to its disappearance does not play a prominent role, otherwise one would, contrary to our data, observe a suppression of the $\Delta^{++}(1232)$ yield in $\nu A$ interactions (respective to $\nu N$ interactions), especially at low momentum region (e.g. $p <$ 1 GeV$/c$), where the absorption cross section via charge-exchange reactions, such as $\Delta^{++} n \rightarrow pp$ and $\Delta^{++} n \rightarrow \Delta^+p$, are expected to be relatively large.

Another characteristic nuclear effect is the baryon production in the kinematically forbidden region (not accessable for interactions on free nucleons), for example, the region of $x_F < - 1$ or the region of $\cos \vartheta_{lab} < 0$, where $\vartheta_{lab}$ is the baryon ejection angle respective to the neutrino direction in the laboratory frame. As it can be seen from Figure 2, practically no $\Delta^{++}(1232)$ production is observed at $x_F < -1$ for $\nu N$ interactions, except a faint yield at $-1.4 < x_F < 0$, reflecting the smearing of the  $x_F$-distribution due to the Fermi-motion of the loosely bound nucleons in the target nucleus. For the case of $\nu A$ interactions, the isobar yield at $x_F < 0$ is not negligible, owing mainly to the intranuclear scattering effects. Further, a possible mechanism of the  $\Delta^{++}(1232)$ ejection to the backward hemisphere ($\cos \vartheta_{lab} < 0$) can be related to the pre-existence of an isobar component in the nuclear wave function \cite{ref19} (see also \cite{ref20} and references therein). As it follows from our data, the $\Delta^{++}(1232)$ production in the backward hemisphere (which corresponds, at our experimental conditions, to the region of $x_F < -1.4$) is negligible, the upper limit of the yield being equal to 0.01 on the 90\% confidence level. This estimation can be compared with the corresponding value of 0.004 inferred in \cite{ref19} from the data on (anti)neutrino-neon interactions at higher energies ($E_\nu = 10-200 GeV$).

\begin{center}
{\large 5. ~SUMMARY}\\
\end{center}

For the first time, the inclusive neutrinoproduction of the $\Delta^{++}(1232)$ isobar from nuclei is investigated. The total yield of $\Delta^{++}(1232)$ is found to be $\langle n\rangle_{\nu A} = 0.080\pm0.011$, being compatible with $\langle n\rangle_{\nu N} = 0.091\pm0.013$ inferred for neutrino-nucleon interactions. Significant nuclear effects are observed in the  $\Delta^{++}(1232)$ inclusive spectra on variables $x_F$, $z$ and $p$, which for $\nu A$ interactions turn out to be shifted towards lower values respective to those for $\nu N$ interactions. This shifting can be caused by the secondary intranuclear collision processes. The $p_T^2$-distributions in $\nu N$ and $\nu A$ interactions are found to be similar, being described with an exponential function with the slope parameter $b(\nu N) = 4.2\pm0.7$ and $b(\nu A) = 4.3\pm1.1$ (GeV$/c)^{-2}$, respectively. No indication on a strong nuclear absorption of the  $\Delta^{++}(1232)$, leading to its disappearance, is found. The $\Delta^{++}(1232)$ production in the backward hemisphere in the laboratory frame is found to be negligible.

\begin{center}
{\large ACKNOWLEDGMENTS}\\
\end{center}

The activity of one of the authors (H.G.) is supported by Cooperation
Agreement between DESY and YerPhI signed on December 6, 2002.


\newpage
\begin{figure}[ht]
\resizebox{0.9\textwidth}{!}{\includegraphics*[bb =20 65 600
610]{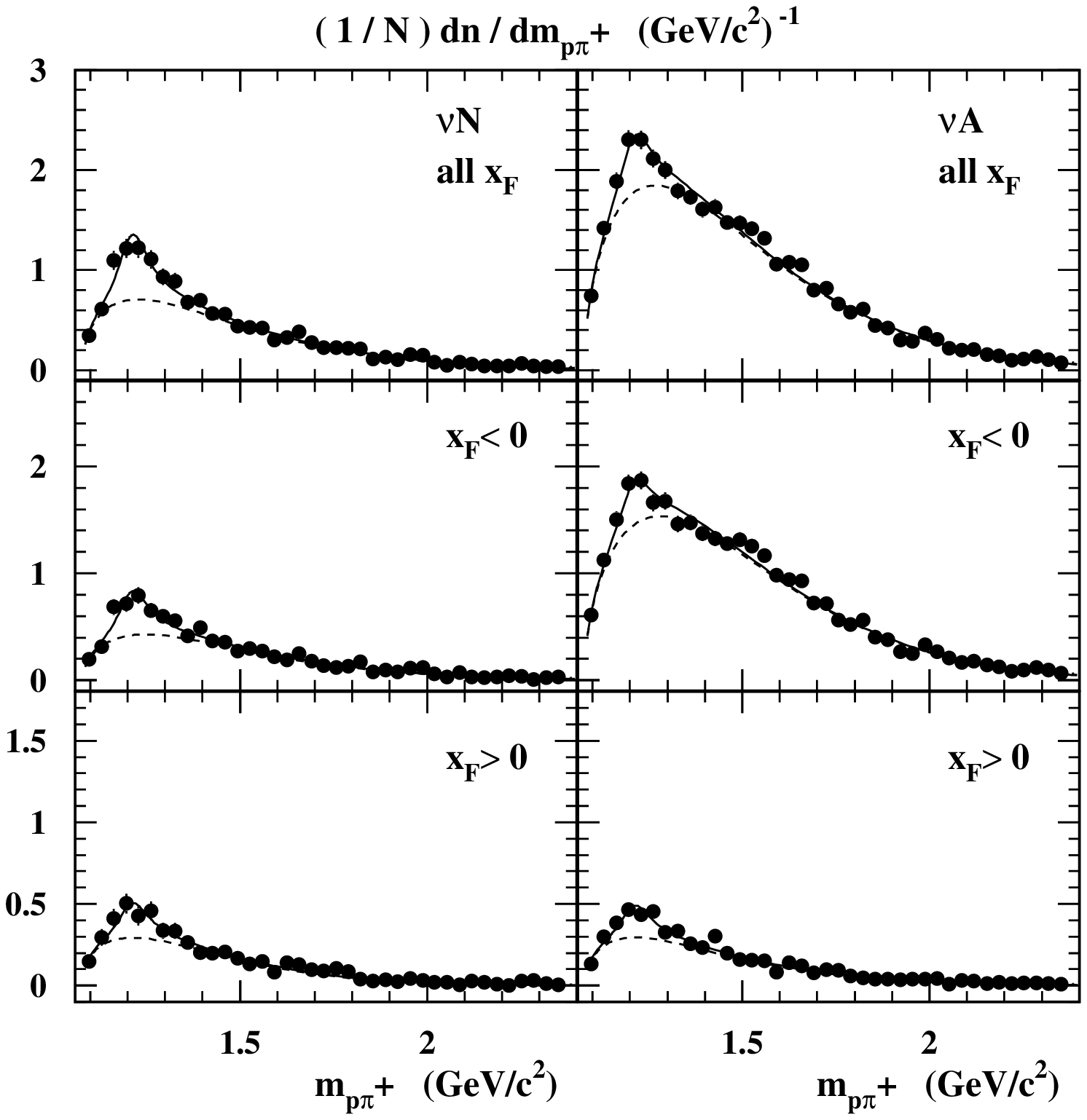}} \caption{The $\pi^+ p$ effective mass
distributions at three ranges of $x_F$ for $\nu N$ (left panel) and $\nu A$ (right panel) interactions. The solid curves are
the fit result, while the dashed curves corresponds to the fitted background distribution (see the text).}
\end{figure}

\newpage
\begin{figure}[ht]
\resizebox{0.9 \textwidth}{!}{\includegraphics*[bb=20 40 500 610]
{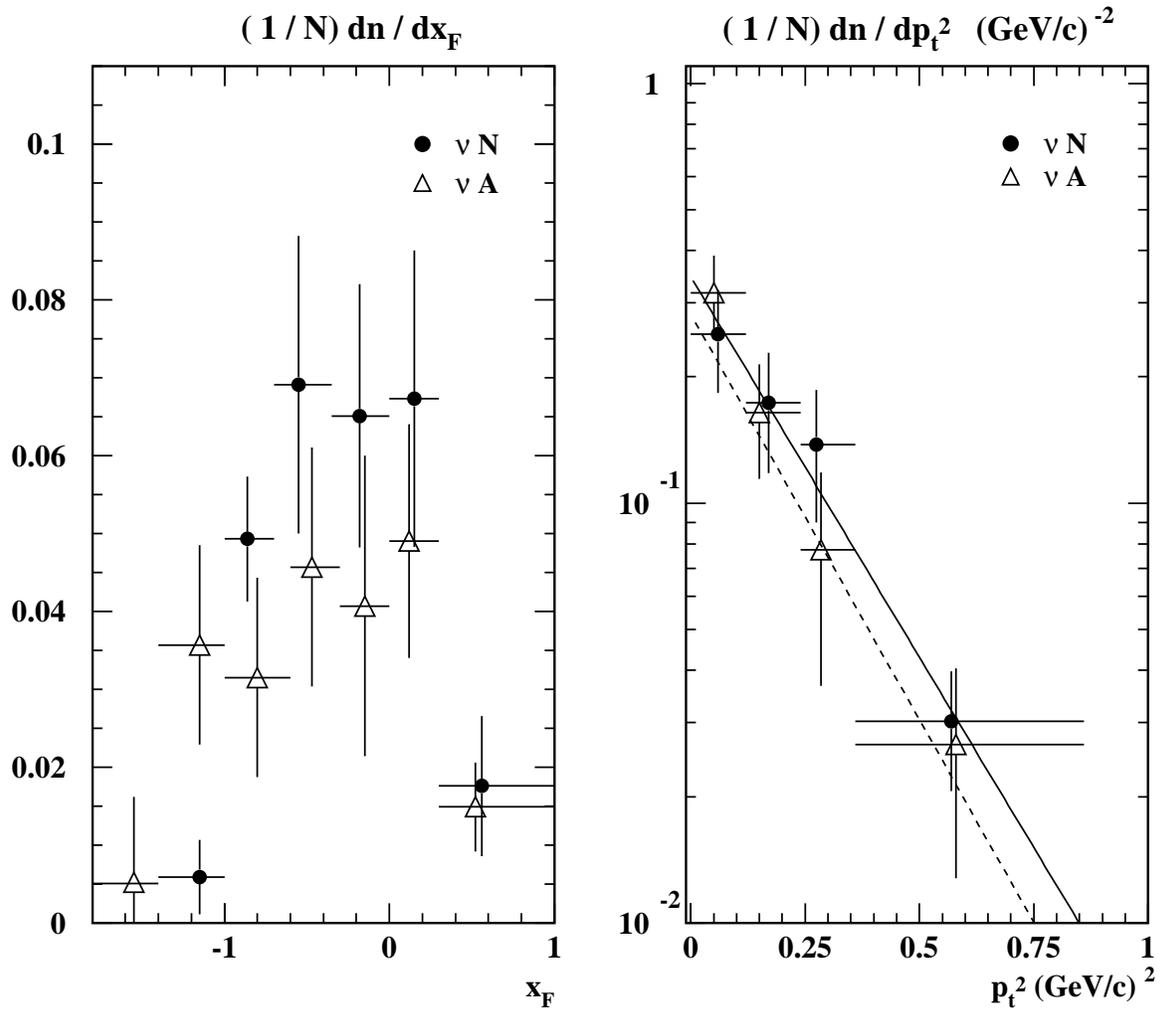}} \caption{The $\Delta^{++}(1232)$ inclusive spectra on $x_F$ (left) and $p_T^2$ (right) in $\nu N$ and $\nu A$ interactions. The solid (dashed) lines are
the fit results for $\nu N$($\nu A)$ interactions.}
\end{figure}

\newpage
\begin{figure}[ht]
\resizebox{0.9\textwidth}{!}{\includegraphics*[bb =20 65 600
610]{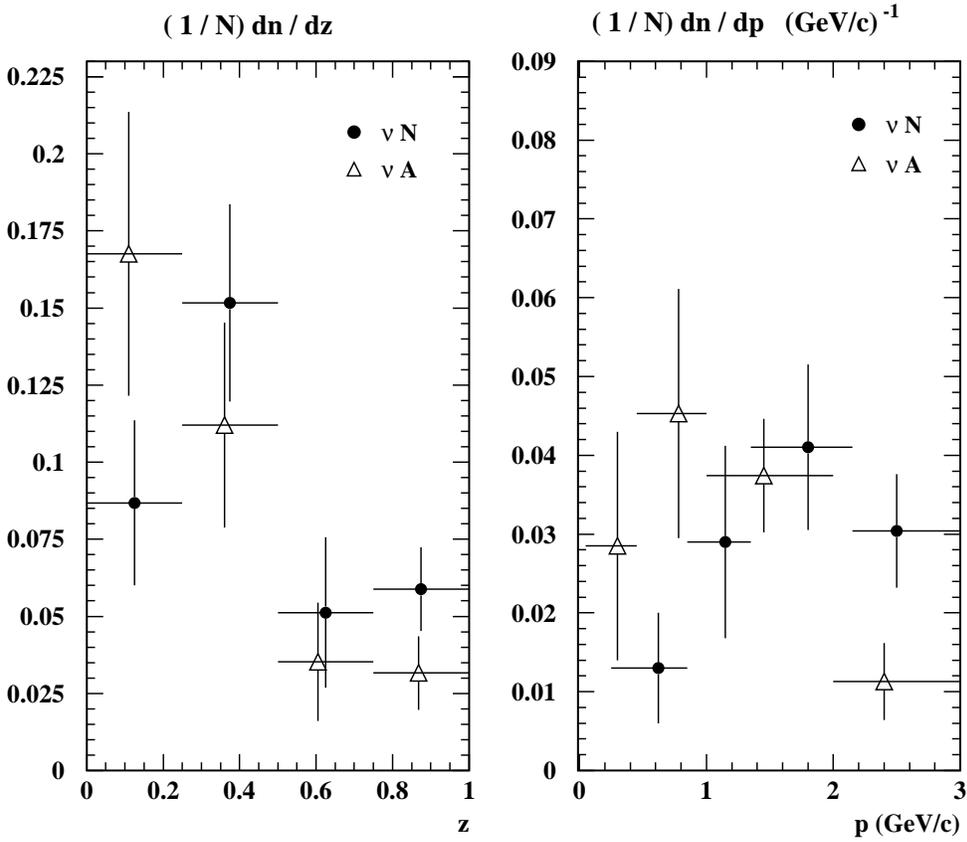}} \caption{The $\Delta^{++}(1232)$ inclusive spectra on the $z$ variable (left) and the total momentum (right).}
\end{figure}

\end{document}